\documentclass[aps,superscriptaddress,floatfix,preprint]{revtex4}
\usepackage{xspace}
\usepackage{graphicx,morefloats}
\usepackage{color}
\usepackage{amsmath, amsthm, amssymb}
\usepackage{multirow}

\graphicspath{{figs/}}
\begin{document}

\title{Quantifying the uncertainties in an ensemble of decadal climate predictions}

\author{Ehud Strobach}
\author{Golan Bel}
\affiliation{Department of Solar Energy and Environmental Physics, Blaustein Institutes for Desert Research, Ben-Gurion University of the Negev, Sede Boqer Campus 84990, Israel}
\begin{abstract}
Meaningful climate predictions must be accompanied by their corresponding range of uncertainty. 
Quantifying the uncertainties is non-trivial, and different methods have been suggested and used in the past.
Here, we propose a method that does not rely on any assumptions regarding the distribution of the ensemble member predictions. 
The method is tested using the CMIP5 1981-2010 decadal predictions and is shown to perform better than two other methods considered here.
The improved estimate of the uncertainties is of great importance for both practical use and for better assessing the significance of the effects seen in theoretical studies.
\end{abstract}

\keywords{Uncertainties, climate predictions, ensemble, decadal climate predictions}

\maketitle

\section{Introduction}
Climate predictions generated by general circulation models should be associated with an estimate for the uncertainties in order to provide meaningful and practical information. 
A common method for assessing the uncertainties in climate predictions is by means of ensembles. The ensemble may consist of different values of the parameters representing their uncertainties, different climate models, different initial conditions, and different initialization methods, as well as any other group of predictions representing the uncertainties in our knowledge of the climate system or the characteristics of the stochastic processes involved.
Often, the ensemble average is used to generate the predictions, and the ensemble spread (the spread is usually quantified either by the standard deviation (STD) for large ensembles or by the entire range spanned by the ensemble predictions for small ensembles) is used to estimate the uncertainties \citep{Deque2007,Smith2013,Strobach_2015b,Woldemeskel2016,Hawkins2016}. In order to consider the ensemble spread as a meaningful estimate for the uncertainties, one must assume a certain probability density function (PDF) of the predicted variable and must consider the predictions of the ensemble members as independent and identically distributed (i.i.d.) variables. These assumptions are not expected to be valid because the different predictions are not truly independent and the real PDF of the variable often deviates from the assumed Gaussian distribution. Moreover, there is no guarantee that the ensemble predictions represent a sample drawn from the actual PDF of the predicted variable. In the context of predictions, the uncertainties are often estimated by the error of the model in predicting past conditions \citep{Deque2007}. The past error is then used to estimate the uncertainties of future predictions. This estimation removes the need for assumptions regarding the distribution of the ensemble predictions, but it still assumes that the errors in the past can serve as a good estimator for future predictions.

Several studies in the past suggested various methods to empirically improve the estimation of the uncertainties.
These methods can be divided into two main groups--regression methods and dressing kernel methods. 
The regression methods ``weight'' the ensemble members in order to minimize a predefined loss function that accounts for the ensemble error and spread. These methods rely on the assumption that the ensemble predictions are i.i.d.  
Common regression methods are the \textit{logistic regression} \citep{Wilks_2006,Wilks_2007} and the \textit{non-homogeneous Gaussian regression} \citep{Gneiting_2005}. 
The dressing kernel methods assume a specific parametric PDF for each of the ensemble members (the characteristics of the PDF are based on the ensemble errors in the past), and the uncertainty is aggregated from the PDFs of all the ensemble members. The dressing kernel methods differ in their assumptions regarding the shape and the width of the ensemble member PDFs. For example, Roulston and Smith \citep{Roulston_2003} assumed Gaussian PDFs and determined their variance from the error spread of the best ensemble member. Wang and Bishop \citep{Wang_2005} determined the variance from the difference between the variance of the ensemble mean errors and the ensemble variance, and Raftery et al. \citep{Raftery_2005} used Bayesian inference to optimize the variance of the dressing kernel. A more thorough  discussion on the various methods can be found in \citep{Wilks_2011}.

Here, we suggest a method for improving the uncertainty estimation by bridging the spread- and error-based estimations. Our method does not rely on any assumption regarding the distribution of the ensemble member predictions or their i.i.d. characteristics. We use the past observed relation between the ensemble mean error and the ensemble spread in order to estimate future prediction uncertainties. Our method relies on the assumption that the relation between the spread and the error does not change significantly between the period used to determine the relation and the period for which the predictions are made. The method was tested on both equally weighted and learning weighted multi-model ensembles of decadal climate predictions \citep{Smith2007,meehl_decadal_2009,meehl_decadal_2014}. It is important to note that decadal climate predictions are based on initialized (using interpolated observed conditions) simulations of climate models \citep{Smith2007,meehl_decadal_2009,meehl_decadal_2014}. The relatively short duration of the simulation (and prediction) period justifies our assumption that the characteristics determined by the results of the initial simulation period will remain valid during the remaining simulation period. We found that the method results in more reliable predictions than those generated by methods that rely on more assumptions.

\section{Methods}
\subsection{Ensemble of climate model predictions}
The ensemble considered here includes simulations of eight climate models from the CMIP5 decadal experiments (see Table \ref{Models_table}). 
We arbitrarily chose one realization of each model. All the models were interpolated to the same spatial resolution of the reanalysis data.
In order to ensure a long enough prediction time series, we focused on the 30-year simulations for the period 1981 to 2011.
The monthly means of the surface temperature were considered as the variable of interest throughout the paper.
The results of a similar analysis for the surface zonal wind appear in the supplementary information.
The first 10 years of the simulations were used to apply the different methods considered here and to tune their parameters while the last 20 years of the predictions were used to validate and measure the performance of the methods.
\begin{table*}
\caption{\label{Models_table} Model Availability}
\centering
\begin{tabular}{ p{3cm}  p{3cm}  p{6cm} p{2.5cm}}
  \hline
  Institute ID  & Model Name & Modeling Center (or Group) & Grid (lat X lon)\\
   \hline
  BCC  &   BCC-CSM1.1  &   Beijing Climate Center, China Meteorological Administration & 64 X 128\\
    CCCma & CanCM4 & Canadian Centre for Climate Modelling and Analysis & 64 X 128\\
    CNRM-CERFACS & CNRM-CM5 & Centre National de Recherches Meteorologiques / Centre Europeen de Recherche et Formation Avancees en Calcul Scientifique & 128 X 256\\
    LASG-IAP  & FGOALS-s2  & LASG, Institute of Atmospheric Physics, Chinese Academy of Sciences & 108 X 128\\
    IPSL* & IPSL-CM5A-LR  & Institute Pierre-Simon Laplace & 96 X 96\\
    MIROC & MIROC5 ~~~~~~~~~~~~~~~ MIROC4h &  Atmosphere and Ocean Research Institute (The University of Tokyo), National Institute for Environmental Studies, and Japan Agency for Marine-Earth Science and Technology & 128 X 256 
    
     320 X 640\\
      MRI & MRI-CGCM3  & Meteorological Research Institute & 160 X 320\\
 \hline
 \end{tabular}
\begin{flushleft}
{\footnotesize * not available for surface zonal wind}
\end{flushleft}
\end{table*}
The NCEP/NCAR reanalysis \citep{kalnay_ncep/ncar_1996} data (with a spatial resolution of $2.5^{\circ}X2.5^{\circ}$) were considered here as observations for the purpose of tuning and validating the methods.
The ERA-interim reanalysis \citep{1321008426928} data were also used, with two different spatial resolutions ($0.75^{\circ}X0.75^{\circ}$ and $2.5^{\circ}X2.5^{\circ}$). We found that the performances of the different methods were not sensitive to the reanalysis data set that we used or to its spatial resolution (obviously, the latter statement is limited to the spatial resolutions that we tested and may not be generally true; see the supplementary information for the results using the ERA-interim reanalysis).
\subsection{Weighted ensemble}
Predictions that are based on an ensemble of models are, in general, a weighted average of the ensemble members. The prediction for time $t$ was defined as:
\begin{equation}
p_t \equiv \sum_{E=1}^N w_{E,t} f_{E,t}.
\end{equation}
Here, $f_{E,t}$ is the prediction of model $E$ for time $t$. $E\in[1..N]$ and $t\in[1..n]$ where $N$ is the number of models in the ensemble and $n$ is the number of time points for which predictions are made. For simplicity, we restricted our attention to the case in which all the models provide forecasts for the same period with equal time steps. 
If there is no a priori knowledge, the weight of all members may be equal. Otherwise, the weight of the models may be based either on previous knowledge or on the past performances of the models.
In order to demonstrate the generality of the methods presented here, we used both an equally weighted ensemble and a weighted ensemble for which the weights were generated by a learning algorithm. 
The learning algorithm that we used is the Exponentiated Gradient Average (EGA)\citep{kivinen_exponentiated_1997,cesa2006prediction} that was shown to outperform the equally weighted ensemble in decadal climate predictions \citep{Strobach_2015,Strobach_2016}.
The learning algorithm used the first 10 years of the simulations to assign weights to the different models, and those weights were then used to generate the predictions for the following 20 years of the simulations. In this case, the weights during the last 20 years were independent of time. It is important to note that the weights were assigned independently for each grid cell.
\subsection{Uncertainty estimation}
One of the main advantages of using an ensemble of models is the fact that one can not only obtain better predictions but also an estimate for the uncertainty range.
The uncertainty range is usually defined as the range within which there is probability $c$ to find the variable. 
We defined the variance of the ensemble prediction at time $t$ as:
\begin{equation}\label{eq:var}
\sigma_t^2 \equiv \sum_{E=1}^N w_E (f_{E,t}-p_t)^2.
\end{equation}
In the equation above and in what follows, we assumed that the weights are independent of time. Because the predictions of different models are not necessarily i.i.d., we still defined the variance of the ensemble according to Eq. \ref{eq:var}, without any prefactor due to the weights of the ensemble members. 

If we assume that the model predictions at each time step are i.i.d. random variables, which is obviously not the case for a multi-model ensemble of climate predictions, then for many models, their average is normally distributed according to the central limit theorem. 
For a standard normal distribution (with a mean equal to zero and an STD equal to one), the interval that includes the variable probability $c$ can be derived from the \textit{probit} function:
\begin{equation}
\delta_G =\sqrt{2} \mathit{erf}^{-1}(c).
\end{equation}
The error function was defined as:
\begin{equation}
\mathit{erf}(s) \equiv \frac{1}{\sqrt{\pi}} \int_{-s}^{s} e^{-x^2} dx.
\end{equation}
Following the above assumptions, we defined the confidence interval for probability $c$ and for time $t$ as:
\begin{equation}
Pr \lbrace \left(p_t - \delta_G \sigma_t\right) \leq y_t \leq \left(p_t + \delta_G \sigma_t\right) \rbrace = c.
\end{equation}
$y_t$ denotes the value of the variable at time $t$. In what follows, we refer to this definition of the confidence interval as the Gaussian estimation.

In most ensembles of climate models, the basic assumptions required for the central limit theorem to hold are not valid and are rarely verified. The number of models in the ensemble is often limited, and the model predictions are not independent and are not necessarily identically distributed. Therefore, the above method for estimating the uncertainty range is not expected to perform well. In order to overcome the abovementioned problems, we suggest here two alternative methods for estimating the uncertainties.

The first method is based on the same assumption of the normal distribution of the ensemble average. Using the ratio between the root mean squared error (RMSE) and the STD of $p_t$ during the learning period (i.e., the first 10 years), a multiplication constant was found to derive a better estimate of the uncertainty range. For an unbiased estimator of $p_t$, the RMSE should be equal to the STD. Therefore, the uncertainty range was derived from a corrected STD, which was defined as $\delta_R\equiv\delta\gamma$ and the correction factor $\gamma$ was defined as:
\begin{equation}
\gamma \equiv \sqrt{\frac{\sum_{t=1}^n (p_t-y_t)^2}{ \sum_{t=1}^n {\sigma_t}^2}} .
\end{equation}
Using this estimation for the uncertainty range, we obtained
\begin{equation}
Pr \lbrace \left(p_t - \delta_R \sigma_t\right) \leq y_t \leq \left(p_t + \delta_R \sigma_t\right) \rbrace = c.
\end{equation}
In what follows, we refer to this uncertainty estimation method as the RMSE-corrected method.
Note that this method is based on the assumption that the ratio between the squared error and the variance does not change much in time and, therefore, can be replaced by its temporal average.

The second method that we introduce here further relaxes the assumption that the prediction is normally distributed by considering an asymmetric uncertainty range around the average. 
In order to define the asymmetric range, we calculated two coefficients: $\gamma_u$ and $\gamma_d$. $\gamma_u$ is a coefficient multiplying the STD of the ensemble in order to set the upper limit of the uncertainty range such that there is a probability of $(1-c)/2$ to find the variable above this upper limit. Similarly, $\gamma_d$ sets the lower limit of the range such that there is a probability $(1-c)/2$ to find the variable below this lower limit. Both coefficients were determined by choosing the minimal range such that during the learning period, the probability of finding the variable between the lower and upper limits is $c$. Mathematically, the coefficients were defined as:
\begin{equation}
\gamma_u=\inf \lbrace \gamma_u \in \Re : \sum_{t=1}^n \Theta((p_t+\gamma_u \sigma_t)-y_t) \geq \frac{1+c}{2} \rbrace,
\end{equation}
and
\begin{equation}
\gamma_d=\inf \lbrace \gamma_b \in \Re : \sum_{t=1}^n \Theta(y_t-(p_t-\gamma_d \sigma_t)) \geq \frac{1+c}{2} \rbrace.
\end{equation}
Using these definitions, the uncertainty range is given by:
\begin{equation}
Pr \lbrace p_t - \gamma_d \sigma_t \leq y_t \leq p_t + \gamma_u \sigma_t \rbrace \approx c
\end{equation}
The non-equality sign here stems from the fact that the probability can change in steps of $1/n$ ($n$ is the number of points in the time series used for calculating $\gamma_u$ and $\gamma_d$). Practically, it implies that for a fine resolution in setting the confidence level $c$, the length of the learning time series, $n$, should be large. In what follows, we refer to this estimation of the uncertainty range as the asymmetric range method. Note that this method defines the range such that upper and lower tails of the distribution have an equal weight of $(1-c)/2$.

It is important to note that all three methods described above base the estimation of the uncertainty range on the STD of the ensemble. Moreover, they all assume that the ratio between the uncertainty range and the STD does not vary much in time. The Gaussian method relies on the assumption of normally distributed model predictions at each time step, the RMSE-corrected method relaxes this assumption and attempts to define a range that is centered on the ensemble prediction, minimizing the prediction bias, and the asymmetric range method attempts to determine a minimal range such that the probability of finding the variable within this range is $c$, regardless of the details of the model predictions distribution.   
Figures 1-5 of the supplementary information present the predictions and their associated uncertainty ranges as estimated by the three methods described above.

The performances of the different methods are measured by the relation between the desired confidence level, $c$, and the actual fraction of the validation time series that fell within the estimated uncertainty range. This fraction is defined as:
\begin{equation}
 F_q=\frac{1}{n_v}\sum_{t=1}^{n_v}\Theta\left(y_t-(p_t-\delta_q\sigma_t)\right)\Theta\left((p_t+\delta_q\sigma_t)-y_t\right),
\end{equation}
where $q$ takes the values $G$ and $R$ for the Gaussian and RMSE-corrected methods, respectively. $n_v$ is the number of time points in the validation series, and the index $t$ measures the time points from the beginning of the validation time series (the period following the learning).
For the asymmetric method, the fraction is defined as:
\begin{equation}
 F_A=\frac{1}{n_v}\sum_{t=1}^{n_v}\Theta\left(y_t-(p_t-\delta_d\sigma_t)\right)\Theta\left((p_t+\delta_u\sigma_t)-y_t\right).
\end{equation}

\section{Results}
We first investigated the ratio between the uncertainty ranges of the surface temperature estimated by the RMSE-corrected and the asymmetric range methods ($ \Delta_R\equiv 2  \delta_R  \sigma_t$ and $\Delta_A\equiv\left(\gamma_u+\gamma_d\right) \sigma_t$, respectively) and those estimated by the Gaussian method ($\Delta_G\equiv2 \delta_G \sigma_t$) for a confidence level of $c=0.9$.
%%%%%%%%%%%%%%%%%%%%%%%%%%%%%%%%%%%
\begin{figure*}[!ht]
\begin{center}
\includegraphics[width=39pc]{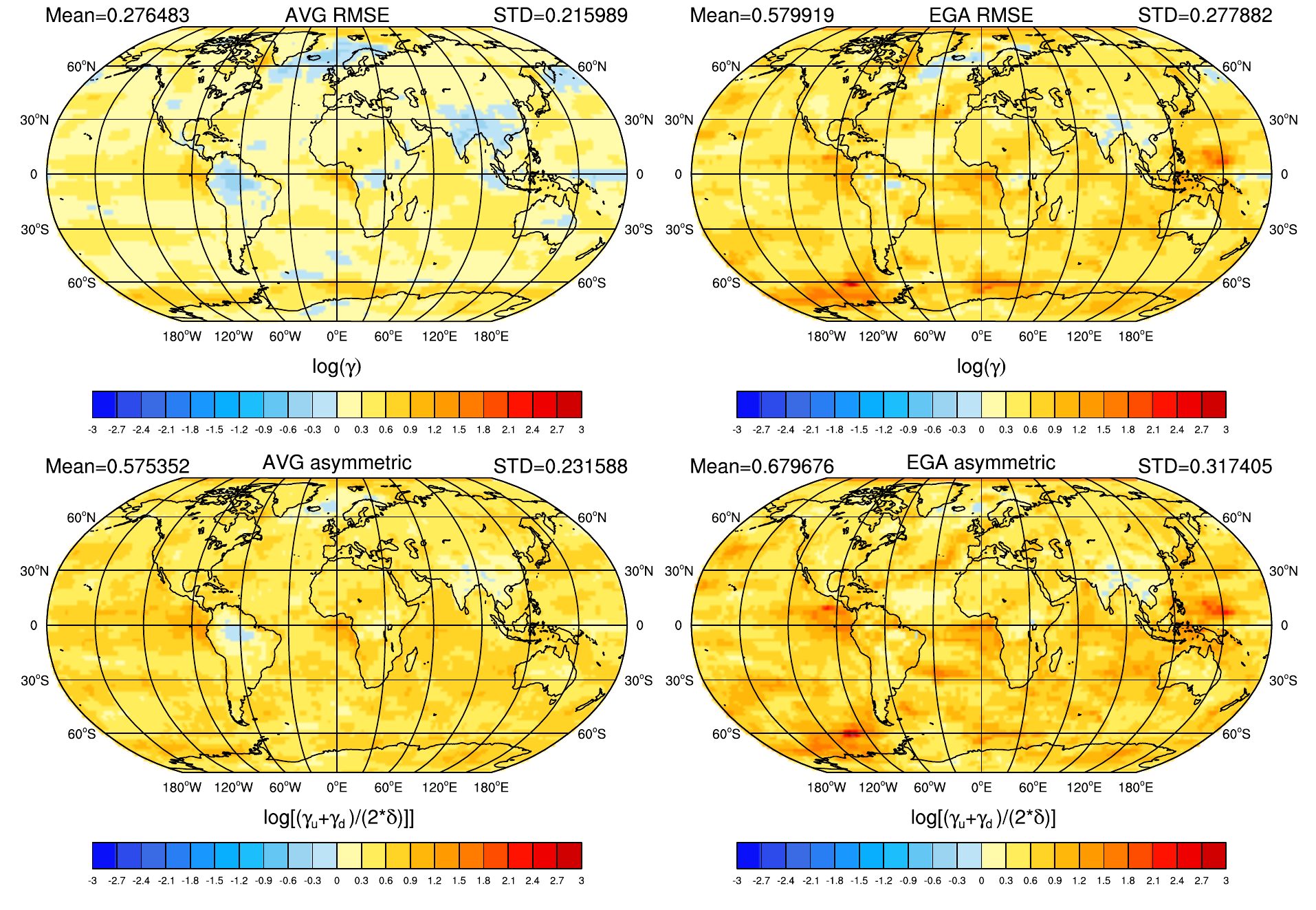} 
\caption{\label{fig:tas_ap=1_gamma-crop} The log of the ratio between the uncertainty ranges of the surface temperature with a confidence level of $0.9$ as estimated by the RMSE-corrected, asymmetric and Gaussian methods. The two left panels show the ratio for the predictions of the equally weighted ensemble, and the two right panels show it for the ensemble weighted by the EGA forecaster. The two upper panels show $\mathit{ln}\left(\Delta_R/\Delta_G\right)$, and the two lower panels show $\mathit{ln}\left(\Delta_A/\Delta_G\right)$. Above each panel, the global mean and the spatial STD are denoted. }
\end{center}
\end{figure*}
%%%%%%%%%%%%%%%%%%%%%%%%%%%%%%%%%%%
Figure \ref{fig:tas_ap=1_gamma-crop} shows the spatial distribution of the natural logarithm of the ratio. The log scale was chosen to emphasize values above and below one (zero in the log scale). Positive values (the ratio is larger than one) correspond to uncertainty ranges (estimated by the RMSE-corrected and the asymmetric range methods) that are larger than the range estimated by the Gaussian method. Similarly, negative values (the ratio is smaller than one) correspond to smaller uncertainty ranges than the range estimated by the Gaussian method. The two left panels show the ratio for the equally weighted ensemble, and the two right panels show it for an ensemble weighted according to the EGA learning algorithm. 
The two upper panels show $\mathit{ln}\left(\Delta_R/\Delta_G\right)$, and the two lower panels show $\mathit{ln}\left(\Delta_A/\Delta_G\right)$. 
Above each panel, the global average (on the left) of the ratio and its STD (on the right) are denoted.

The upper panels of Fig. \ref{fig:tas_ap=1_gamma-crop} show that, in general, our ensemble forecast is overconfident, i.e., the spread of the ensemble predictions is smaller than the typical error, and it becomes even more overconfident when the weighting is done according to the EGA learning algorithm. The spatial variability of the ratio is not very large. 

The performances of the different methods of estimating the uncertainty range can be tested by comparing the desired confidence, $c$, to the fraction of the observations within the predicted range, $F_q$.
$F_q<c$ implies overconfidence (i.e., too narrow a range) and vice versa.
%%%%%%%%%%%%%%%%%%%%%%%%%%%%%%%%%%%%%
\begin{figure*}[!ht]
\begin{center}
\includegraphics[width=39pc]{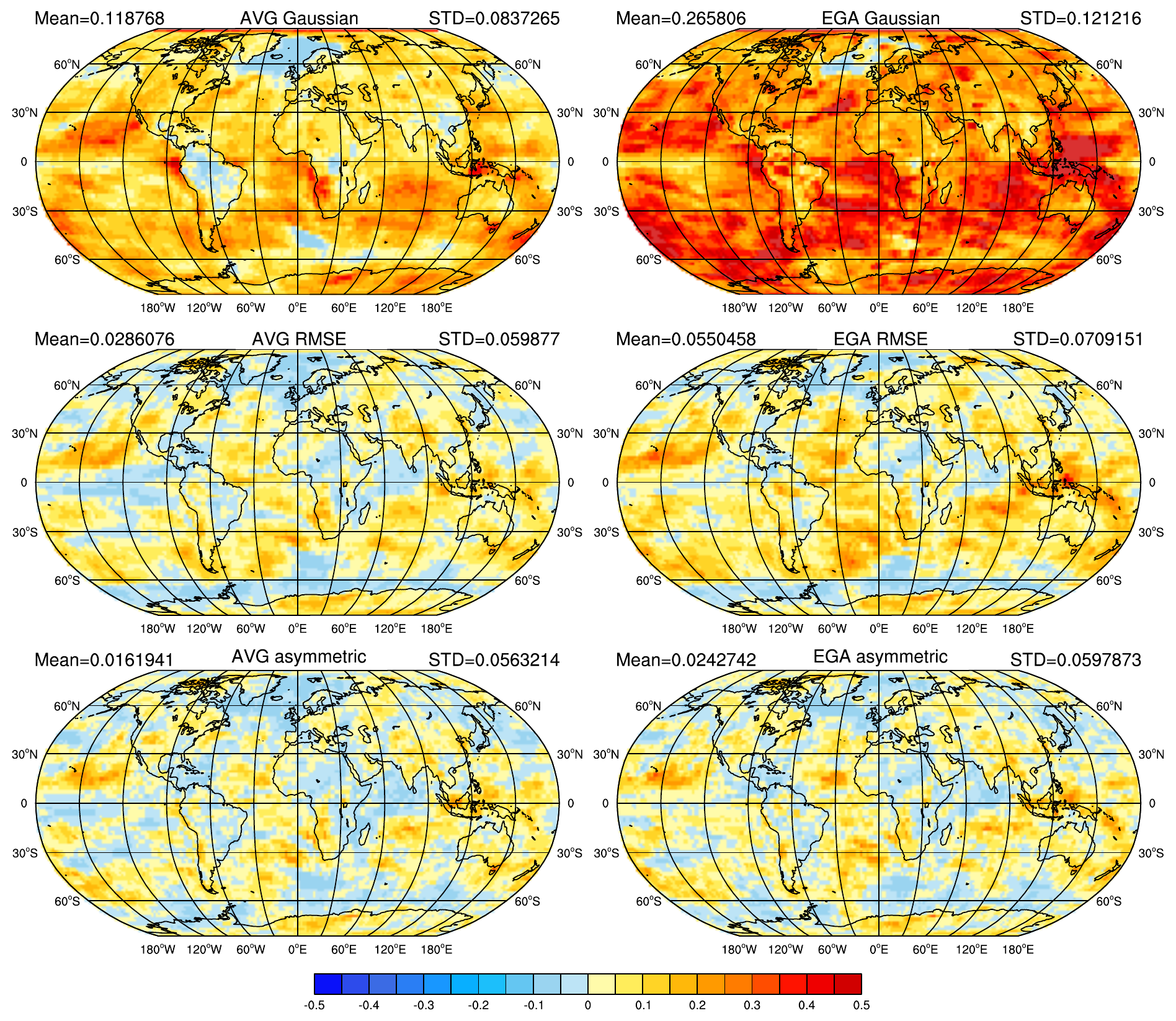} 
\caption{\label{fig:tas_ap=1_rel_diagram_global9-crop} The spatial distribution of the difference between the fraction of the observations that were outside the predicted range of the $c=0.9$ confidence level and the predicted $0.1$ fraction ($c-F_q$), for the surface temperature. The three left panels show the difference for the equally weighted ensemble, and the three right panels show it for the EGA forecaster. 
The two upper panels show ($c-F_G$), the two center panels show ($c-F_R$), and the two lower panels show ($c-F_A$). The spatial average and the STD of the difference are denoted above each panel.
}
\end{center}
\end{figure*}
%%%%%%%%%%%%%%%%%%%%%%%%%%%%%%%%%%%%%%%%%%%%%%%%%%%%%%%%%%
Figure \ref{fig:tas_ap=1_rel_diagram_global9-crop} shows the difference between the fraction of the observations that were outside the predicted range of the $0.9$ confidence level and the predicted $0.1$ fraction ($c-F_q$). 
The three left panels show the difference for the equally weighted ensemble, and the three right panels show it for the EGA forecaster. 
The two upper panels show $c-F_G$, the two center panels show $c-F_R$, and the two lower panels show $c-F_A$.
Above each panel, we denote the spatial average (on the left) and the STD (on the right) of the difference.
The figure shows that in most regions, $\Delta_G$ is too small ($c>F_G$), which justifies the larger values of $\Delta_R$ and $\Delta_A$. The underestimation of $\Delta$ by $\Delta_G$ is more severe for the EGA than for the equally weighted ensemble. 

Figure \ref{fig:tas_ap=1_rel_diagram_global9-crop} demonstrates that the RMSE-corrected and asymmetric methods outperform the Gaussian method. Moreover, it shows that the asymmetric method outperforms the RMSE-corrected method. One can also notice that the methods work better for the equally weighted ensemble than for the EGA forecaster. This better performance is due to the fact that the ensemble STD, $\sigma_t$, of the EGA is defined by weights that minimize the forecast error and not by weights that maximize the forecast reliability. 
In the supplementary information, we provide similar results for the surface zonal wind (Fig. 7) and for the temporal evolution of $F_q$ (Figs. 8-9). The results accord with the results presented here for the surface temperature and with the basic assumptions of the methods for estimating the uncertainty range.
%%%%%%%%%%%%%%%%%%%%%%%%%%%%%%%%%%%%%%%%%%%%%%
\begin{figure*}[!ht]
\begin{center}
\includegraphics[width=27pc]{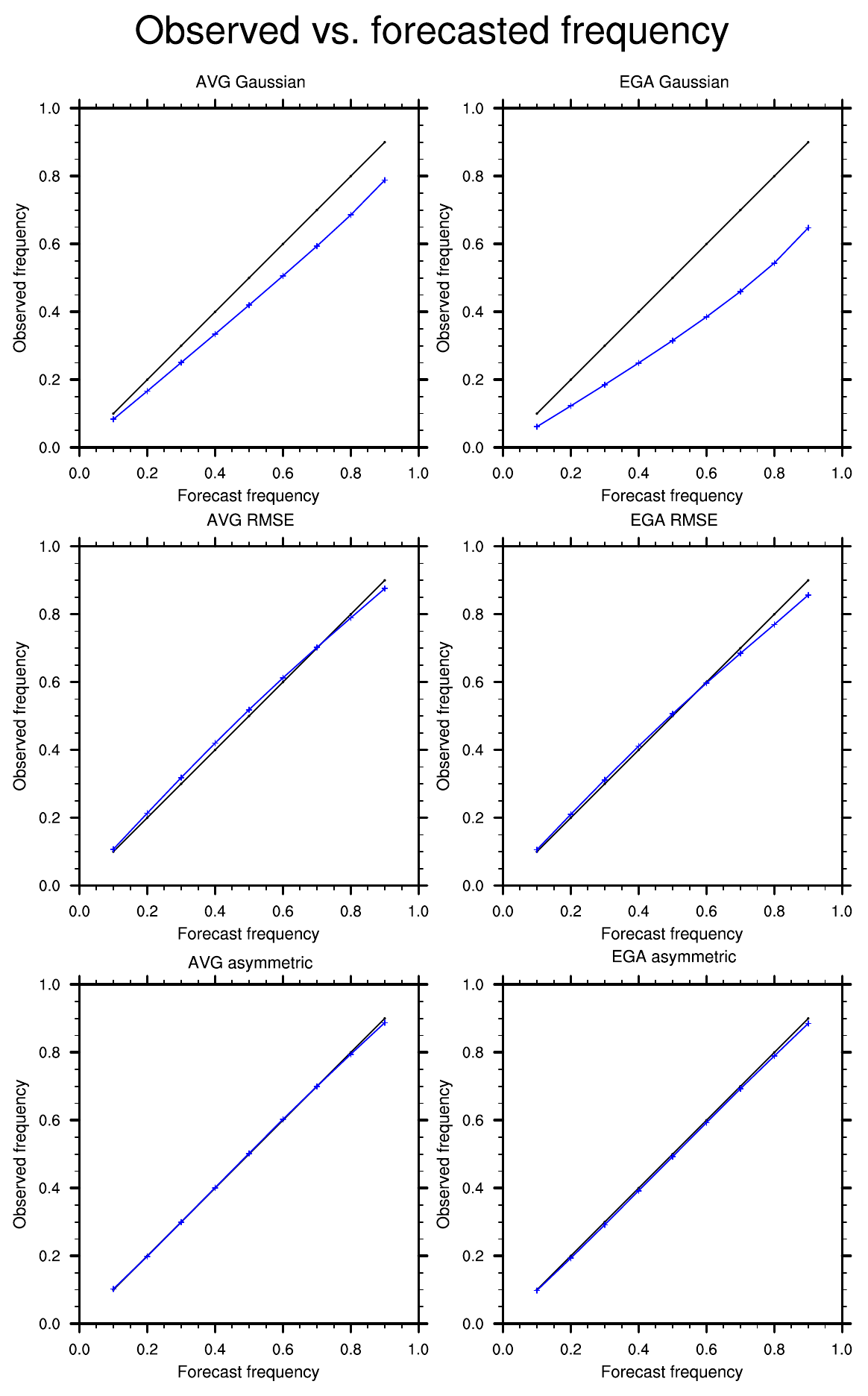} 
\caption{\label{fig:tas_ap=1_rel_diagram-crop}  Reliability diagrams for the different estimation methods of the surface temperature uncertainty range.
The observed frequency represents the spatial average over all the grid cells. The left panels show the reliability for the equally weighted ensemble, and the right panels show it for the EGA forecaster. 
The upper panels show the reliability of the Gaussian method, the center panels show the reliability of the RMSE-corrected method, and the lower panels show the reliability of the asymmetric method.}
\end{center}
\end{figure*}
%%%%%%%%%%%%%%%%%%%%%%%%%%%%%%%%%%%%%%%%%%
These methods can be used for different confidence levels. Tests of the method performances that span the whole range  of confidence levels, $c=[0,1]$, provide more information about the relation between the estimated and observed distributions of the variable. 
Figure \ref{fig:tas_ap=1_rel_diagram-crop} shows the observed frequency versus the expected one (reliability diagrams) for different confidence levels.
The observed frequencies represent the spatial average over the whole globe.
The left (right) panels show the reliability diagrams for the equally weighted ensemble (EGA forecaster). 
The upper panels show the reliability for the Gaussian method, the middle panels show the reliability for the RMSE-corrected method, and the lower panels show the reliability for the asymmetric method.

Figure \ref{fig:tas_ap=1_rel_diagram-crop} shows that the RMSE-corrected and asymmetric methods perform much better than the Gaussian method. The improvement is more pronounced for the EGA forecaster. One can also observe that the asymmetric method is more reliable for higher confidence levels.
Similar results were obtained for the surface zonal wind. However, we found that in most cases, the uncertainty range for the surface zonal wind was underestimated by all the methods (see Fig. 8 of the supplementary information).

\section{Summary and Discussion}

A new method for estimating the uncertainty of ensemble-based climate predictions was suggested.
This method is based on learning the relations between the prediction errors and their spread in the past and using these relations in order to estimate the uncertainties of future predictions for which only the spread is known. 
The method does not rely on any assumptions regarding the distribution of the ensemble member predictions or that they are i.i.d. 
It also has the advantage of estimating separately the upper and lower uncertainties.
The inherent assumption of the method is that the relation between the spread of the ensemble predictions and the errors does not significantly change during the period spanned by the predictions. Moreover, it assumes that this relation has relatively small fluctuations, which allows it to be considered as independent of time.

The excellent performance of the method in estimating the uncertainties during a validation period, based on the relations found during an earlier learning period, suggests that these assumptions are valid for a multi-model ensemble of decadal climate predictions. Moreover, it was shown that the method performs well for both the equally weighted ensemble and the ensemble weighted according to a learning algorithm applied during the training period. The performance of the asymmetric method was better than those of the RMSE-corrected method and the method based on the assumption of a Gaussian distribution of the ensemble member predictions. 
It was also found that, in general, estimations of the uncertainties that are based on the ensemble spread without correction result in overconfident predictions (therefore, the correction factors are larger than one). The overconfidence is even stronger for the weighted ensemble. It is possible that weighting schemes that are based on the reliability of the predictions and not just on their error may result in a spread that is closer to the errors; however, a deeper discussion of this issue is beyond the scope of this paper.
The method suggested here is not limited to estimating one measure of the uncertainties (for example, the variance); applying the method with different confidence intervals was shown to provide an excellent estimation of the whole PDF of the variables.
The asymmetric range method is not limited to decadal climate predictions and is likely to be useful for climate predictions of shorter time scales or for weather predictions.
In particular, the method may be useful for estimating the uncertainties for data assimilation in numerical weather predictions \citep{Houtekamer1996,Kalnay2003,Warner2010}.
 
\clearpage

\acknowledgments
The research leading to these results has received funding from the European Union Seventh Framework Programme (FP7/2007-2013) under grant number [293825].

We acknowledge the World Climate Research Programme's Working Group on Coupled Modelling, which is responsible for the CMIP, and we thank the climate modeling groups (listed in Table \ref{Models_table} of this paper) for producing and making available their model outputs. For the CMIP, the U.S. Department of Energy's Program for Climate Model Diagnosis and Intercomparison provides coordinating support and leads the development of software infrastructure in partnership with the Global Organization for Earth System Science Portals.

\bibliographystyle{unsrt}
\bibliography{references}

\appendix
\section{Supplementary Information}

In this supplementary information, we provide: i) examples of time series of predictions and their confidence intervals; ii) results for the uncertainty estimates of the surface zonal wind; iii) time series showing the fraction of grid cells for which the observations are outside the predicted range; iv) results for estimating the uncertainties using the ERA-interim reanalysis data; and v) diagrams showing the temporal and spatial averages of the predicted ranges versus the actual ranges.

\subsection{Time series}

\begin{figure*}
\begin{center}
\includegraphics[width=29pc]{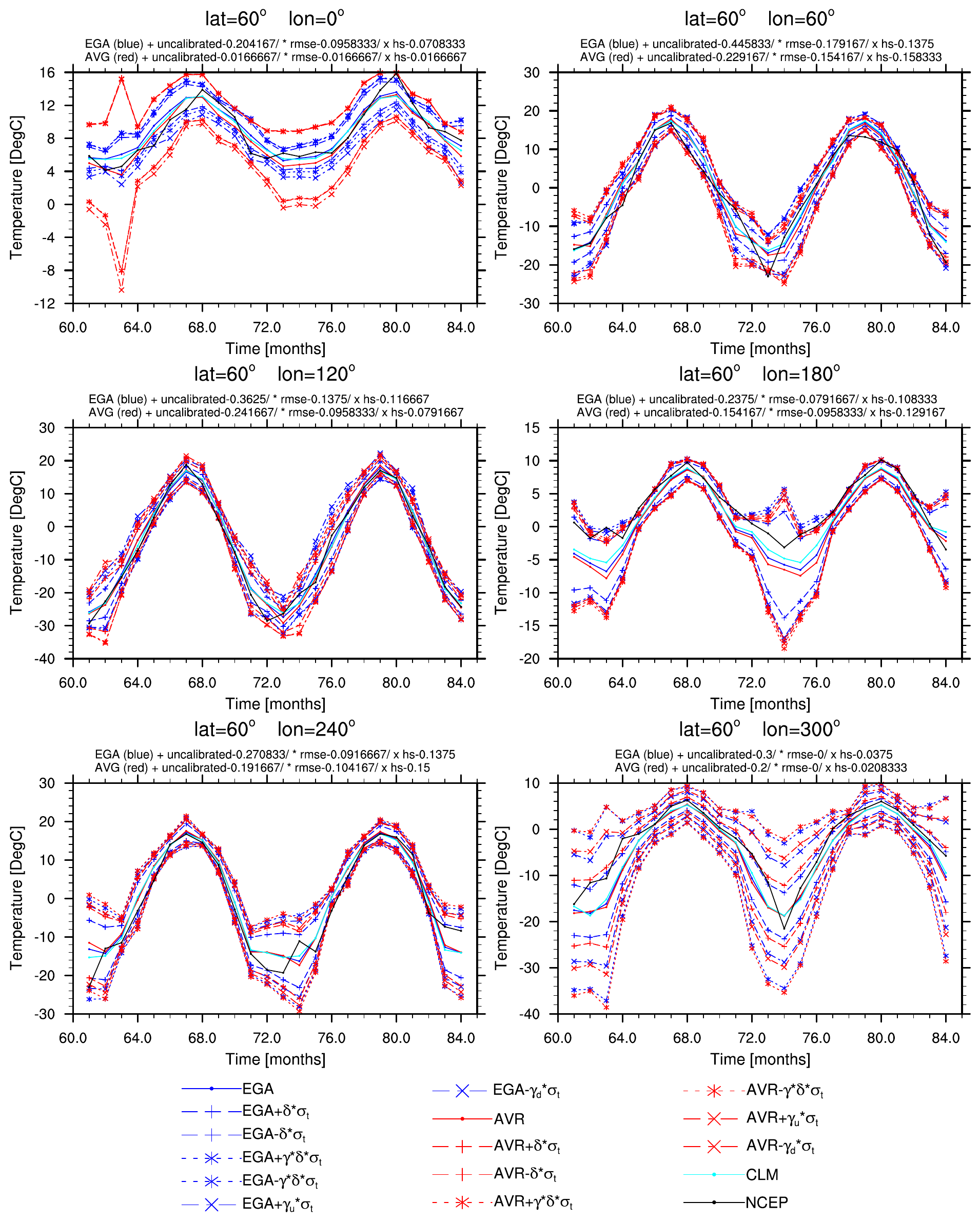} 
\caption{\label{fig:tas_1_5-crop} Time series of the monthly averages of surface temperature for the observations, climatology, and predictions of the simple average and the EGA forecasters (two years from twenty years of predictions). The $90\%$ confidence interval is indicated based on the three methods described in the main text. Above each panel, the fraction of the observations outside the predicted ranges is indicated (for the twenty-year time series). $lat = 60^\circ$ and the different $lon$ values are indicated above each panel. }
\end{center}
\end{figure*}

\begin{figure*}
\begin{center}
\includegraphics[width=29pc]{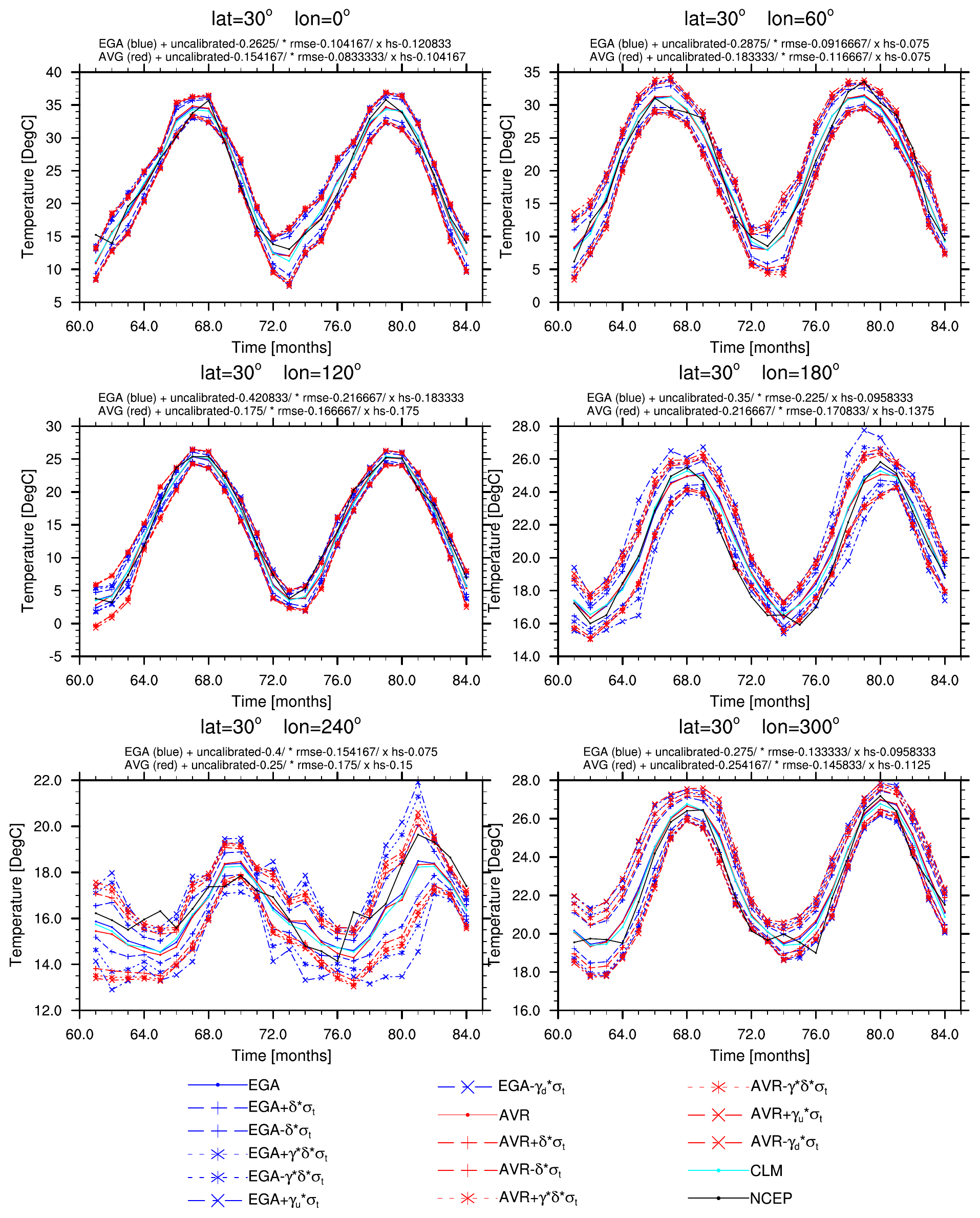} 
\caption{\label{fig:tas_1_4-crop} Similar to Fig. \ref{fig:tas_1_5-crop} but for $lat = 30^\circ$.
}
\end{center}
\end{figure*}

\begin{figure*}
\begin{center}
\includegraphics[width=29pc]{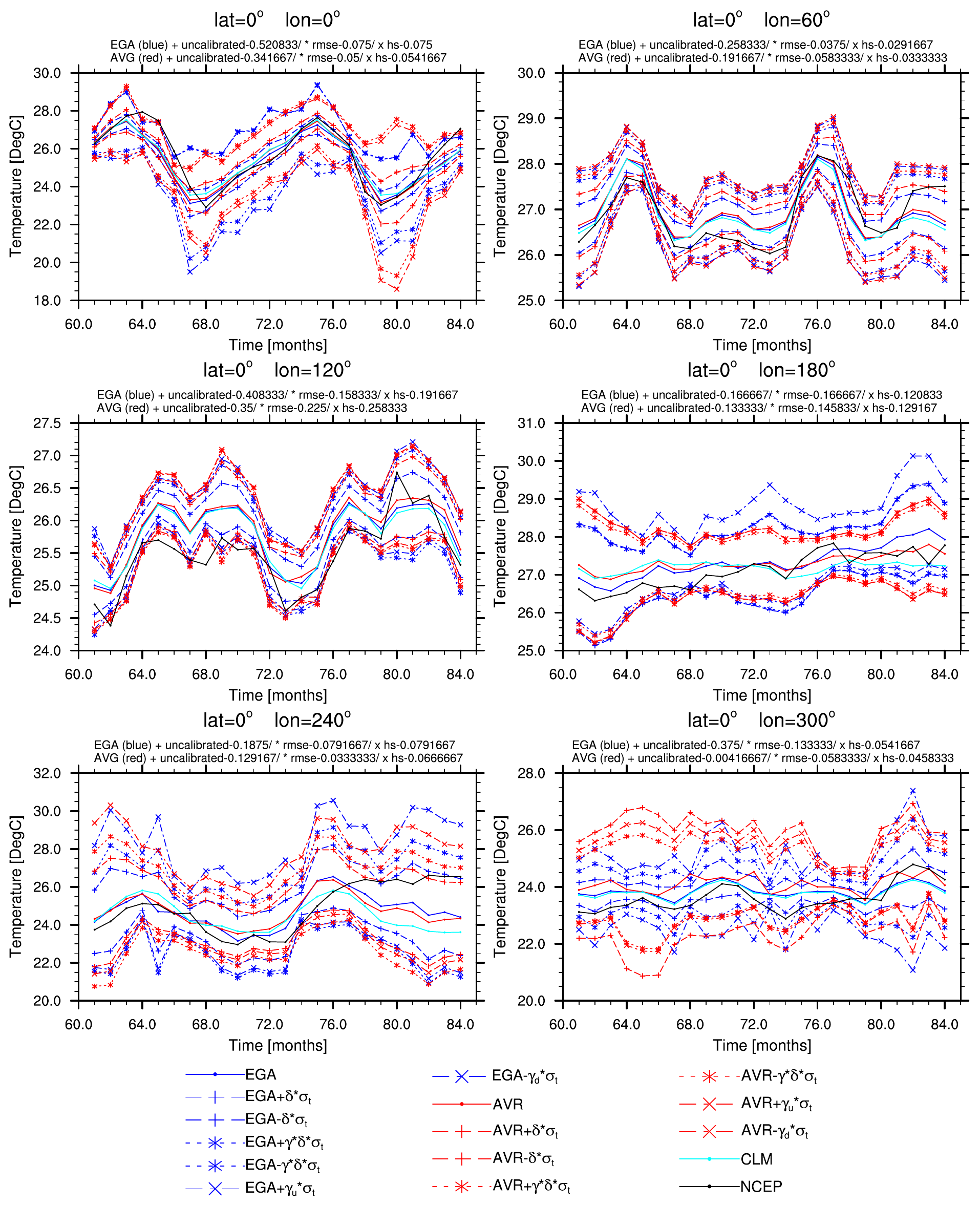} 
\caption{\label{fig:tas_1_3-crop} 
Similar to Fig. \ref{fig:tas_1_5-crop} but for $lat = 0^\circ$.
}
\end{center}
\end{figure*}

\begin{figure*}
\begin{center}
\includegraphics[width=29pc]{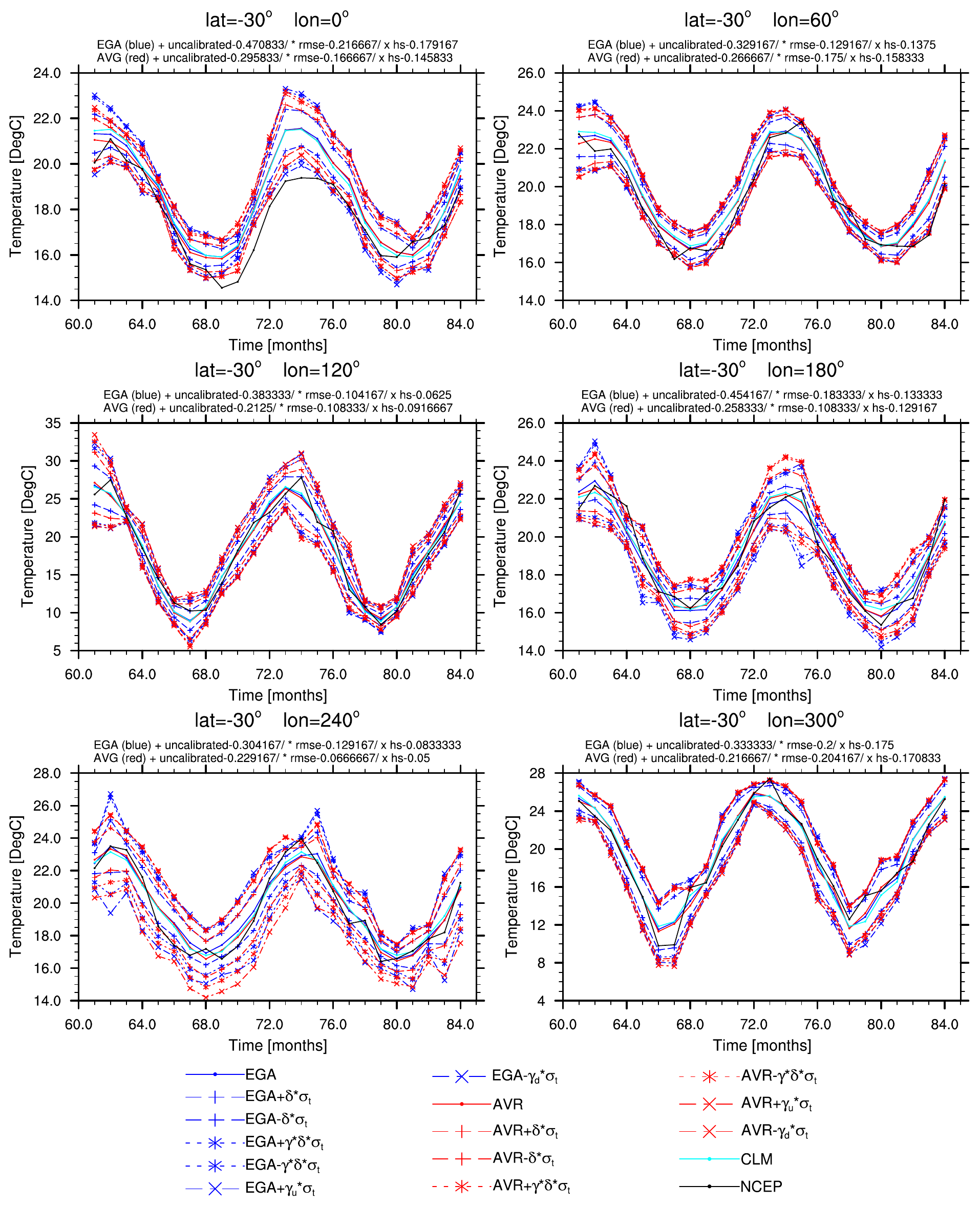} 
\caption{\label{fig:tas_1_2-crop} 
Similar to Fig. \ref{fig:tas_1_5-crop} but for $lat = -30^\circ$.
}
\end{center}
\end{figure*}

\begin{figure*}
\begin{center}
\includegraphics[width=29pc]{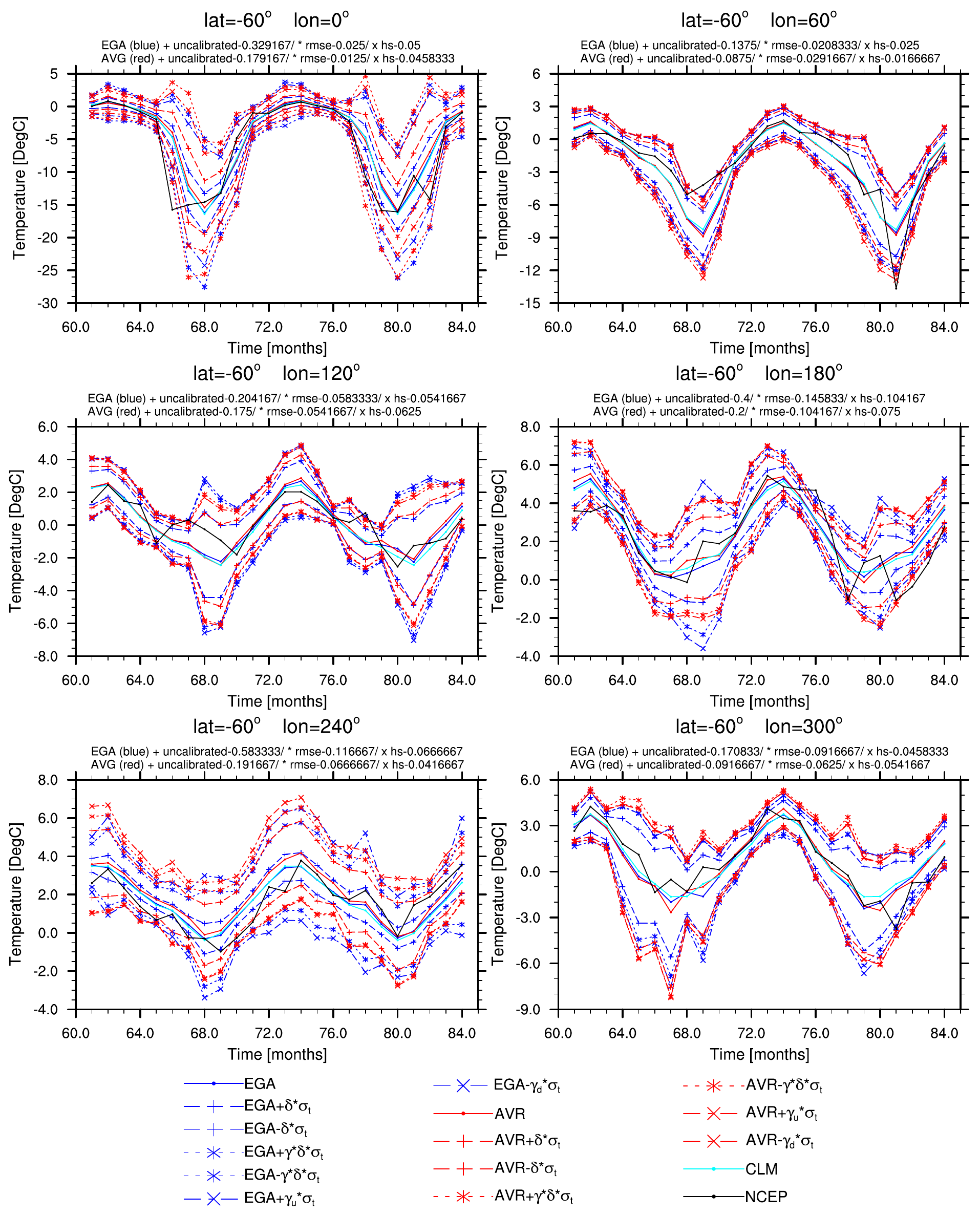} 
\caption{\label{fig:tas_1_1-crop} 
Similar to Fig. \ref{fig:tas_1_5-crop} but for $lat = -60^\circ$.
}
\end{center}
\end{figure*}

Figures \ref{fig:tas_1_5-crop}-\ref{fig:tas_1_1-crop} show examples of time series of predictions and their corresponding $90\%$ confidence intervals based on the three methods that were considered in the main paper. The predictions made by the simple average and the EGA are displayed and compared with the climatology and the NCEP reanalysis.

%\clearpage

\subsection{Surface zonal wind}

\begin{figure*}
\begin{center}
\includegraphics[width=29pc]{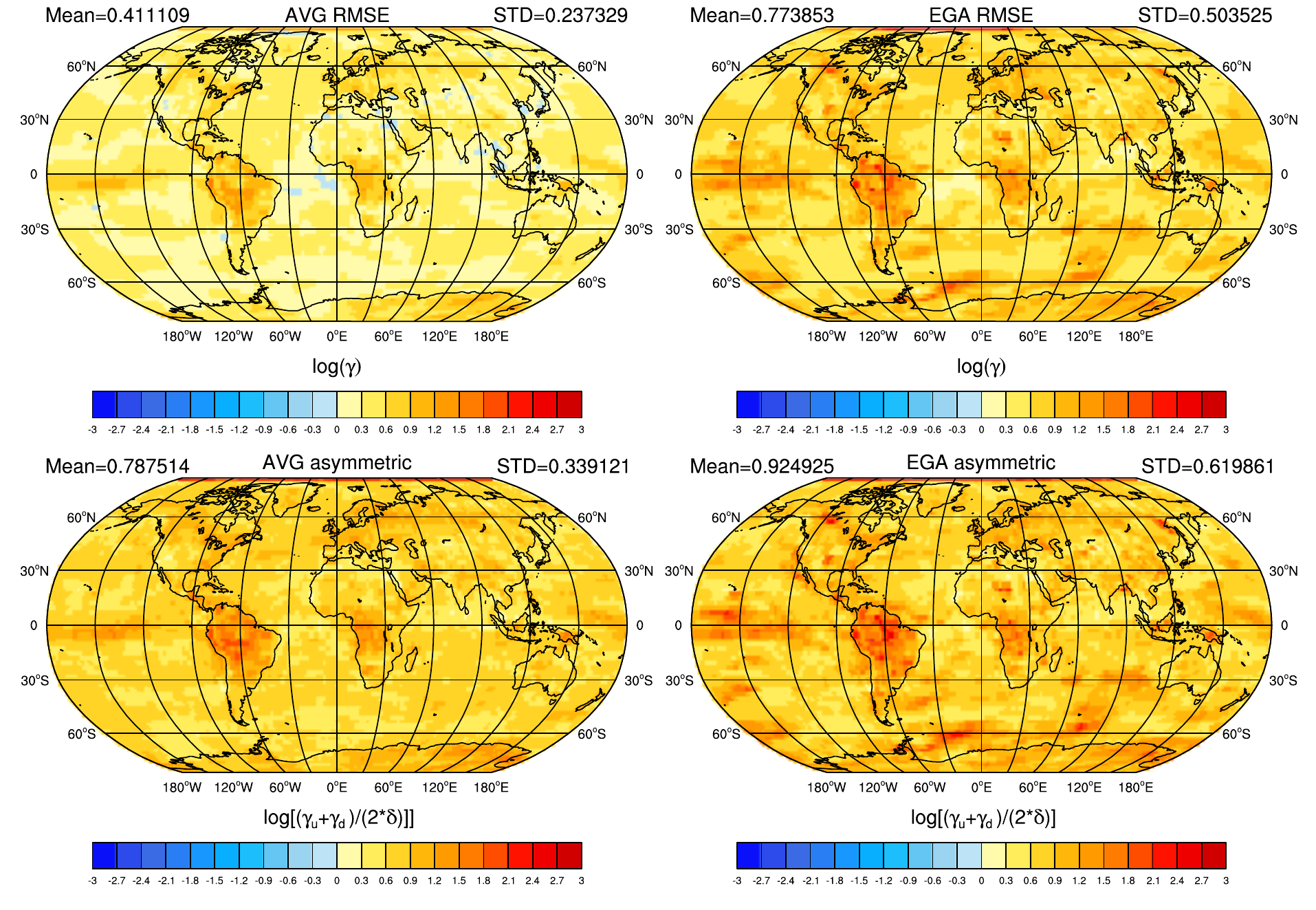} 
\caption{\label{fig:uas_ap=1_gamma_log-crop} The log of the ratio between the uncertainty ranges of the surface zonal wind with a confidence level of 0.9 as estimated by the RMSE-corrected and asymmetric methods and those estimated by the Gaussian method. The two left panels show the ratio for the predictions of the equally weighted ensemble, and the two right panels show it for an ensemble weighted by the EGA forecaster. The two upper panels show $\ln(\Delta_R /\Delta_G )$, and the two lower panels show $\ln(\Delta_A /\Delta_G )$. Above each panel, the global mean and the spatial STD are denoted. }
\end{center}
\end{figure*}

\begin{figure*}
\begin{center}
\includegraphics[width=29pc]{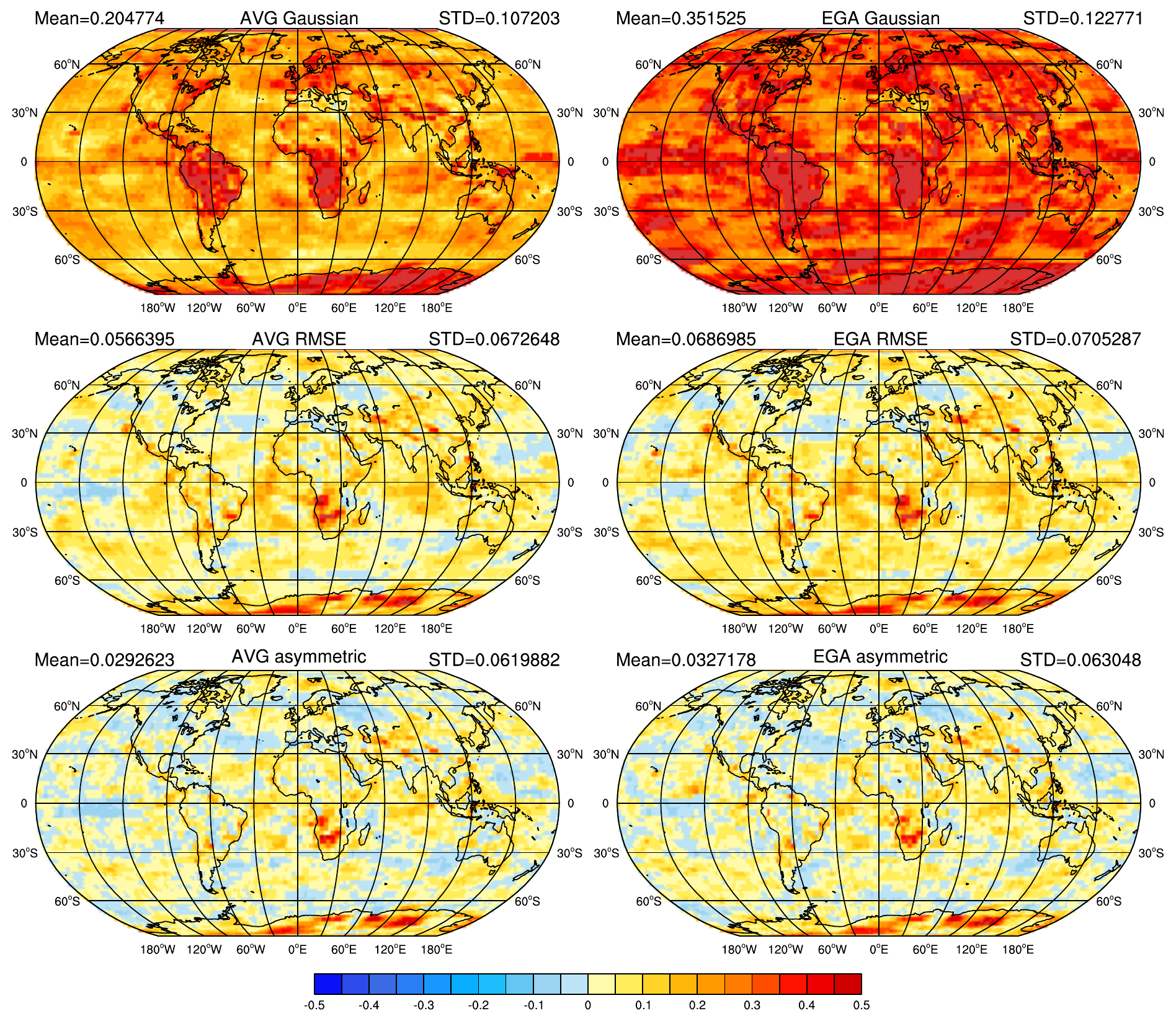} 
\caption{\label{fig:uas_ap=1_rel_diagram_global9-crop} The spatial distribution of the difference between the fraction of the observations that were outside the predicted range of the $c=0.9$ confidence level and the predicted $0.1$ fraction ($c-F_q$), for the surface zonal wind. The three left panels show the difference for the equally weighted ensemble, and the three right panels show it for the EGA forecaster. 
The two upper panels show ($c-F_G$), the two center panels show ($c-F_R$), and the two lower panels show ($c-F_A$). The spatial average and the STD of the difference are denoted above each panel.
}
\end{center}
\end{figure*}

\begin{figure*}
\begin{center}
\includegraphics[width=25pc]{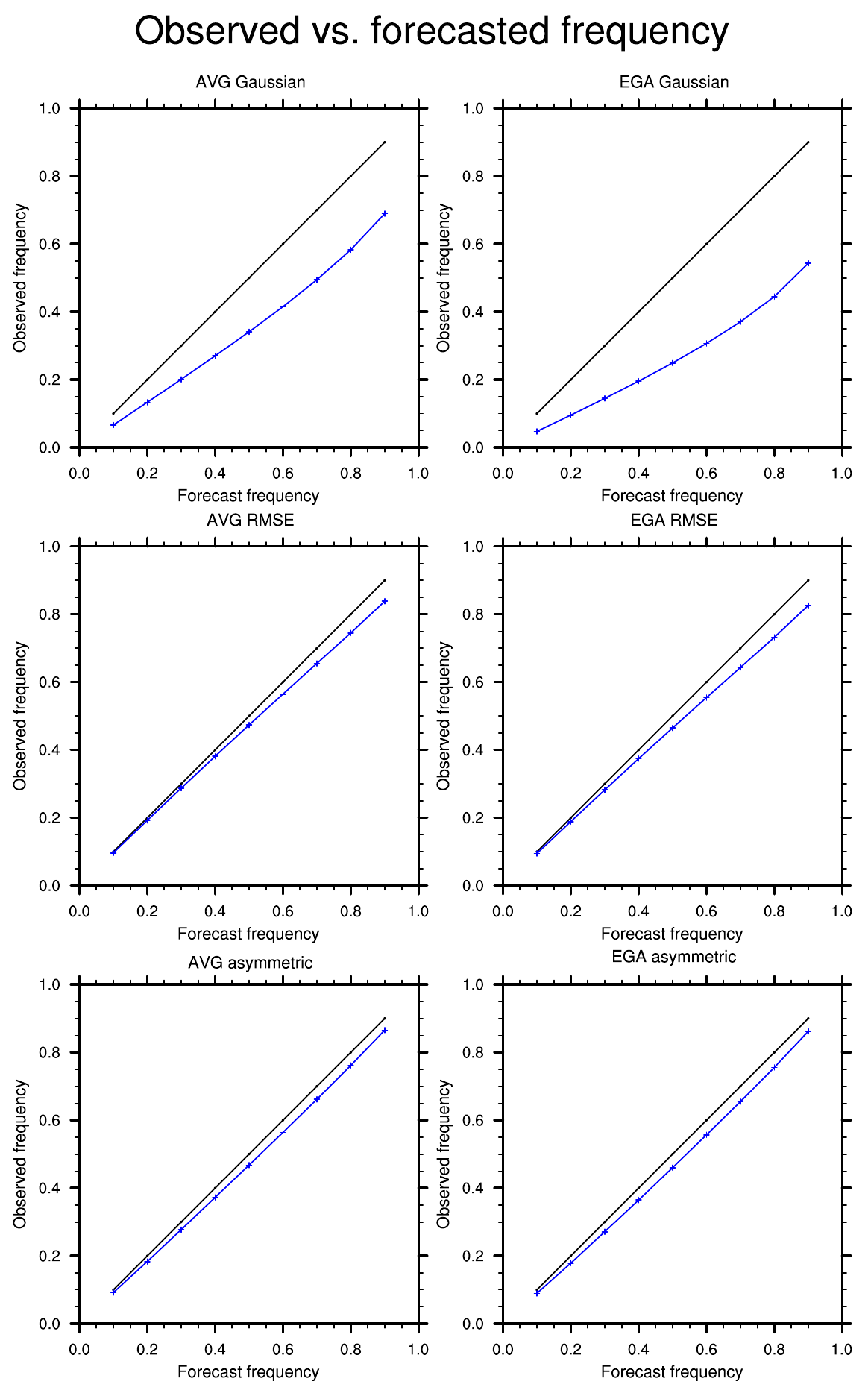} 
\caption{\label{fig:uas_ap=1_rel_diagram-crop} Reliability diagrams for the different estimation methods of the surface zonal wind uncertainty range.
The observed frequency represents the spatial average over all the grid cells. The left panels show the reliability for the equally weighted ensemble, and the right panels show it for the EGA forecaster. 
The upper panels show the reliability of the Gaussian methods, the center panels show the reliability of the RMSE-corrected method, and the lower panels show the reliability of the asymmetric method. } \end{center}
\end{figure*}

Figures \ref{fig:uas_ap=1_gamma_log-crop}-\ref{fig:uas_ap=1_rel_diagram-crop} show similar information to that presented in Figs. 1-3 in the main paper, but for the surface zonal wind.

%\clearpage

%\clearpage

\subsection{Global time series}

\begin{figure*}
\begin{center}
\includegraphics[width=29pc]{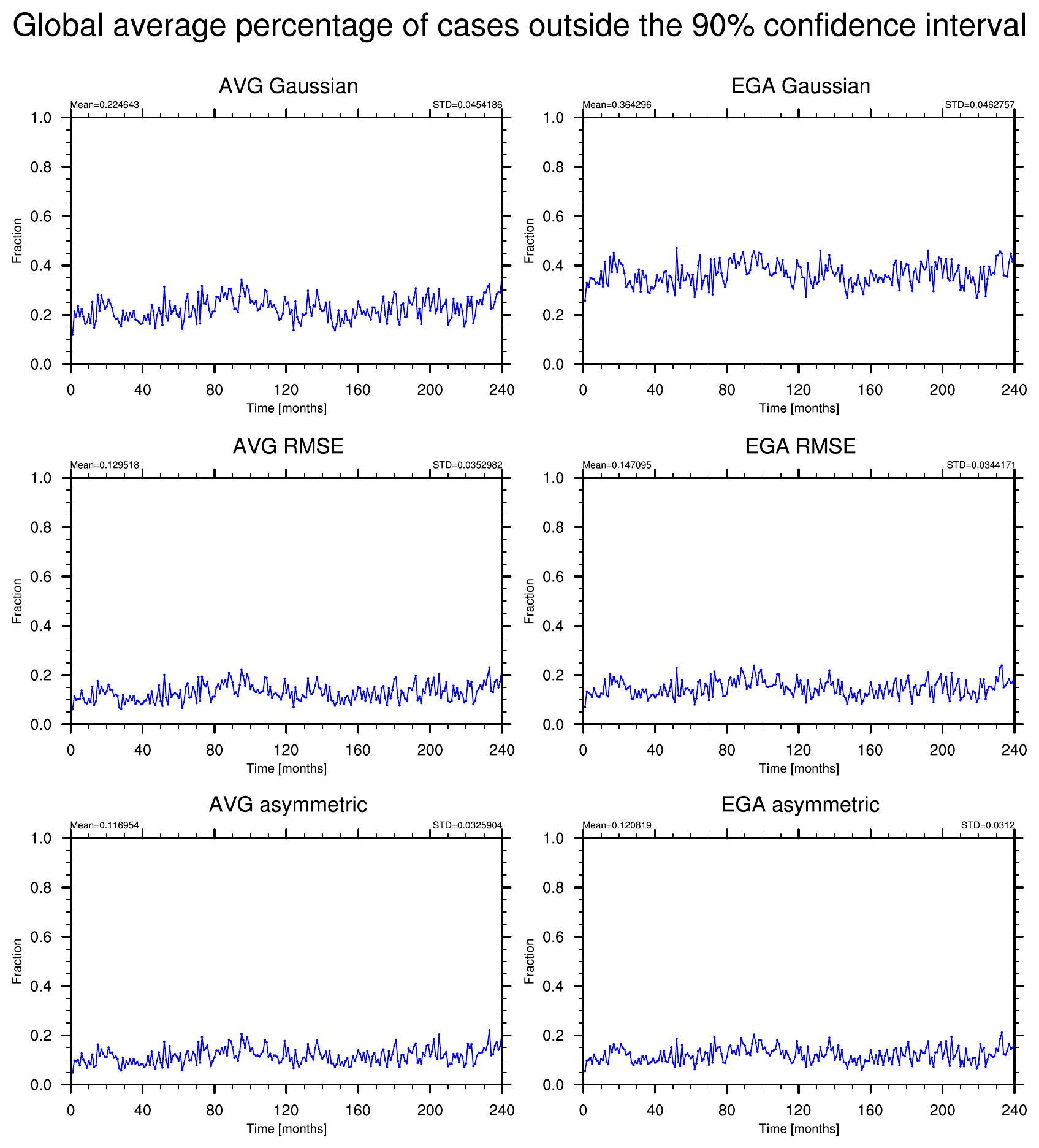} 
\caption{\label{fig:tas_ap=1_rel_year-crop} The fraction of the observations that were outside the predicted range of the $0.9$ confidence level for surface temperature. The three left panels show the fraction for the simple average, and the three right panels show it for the EGA forecaster. The two upper panels show $(c - F_G )$, the two center panels show $(c - F_R)$, and the two lower panels show $(c - F_A)$. The spatial average and the STD of the difference are denoted above each panel.}
\end{center}
\end{figure*}

\begin{figure*}
\begin{center}
\includegraphics[width=29pc]{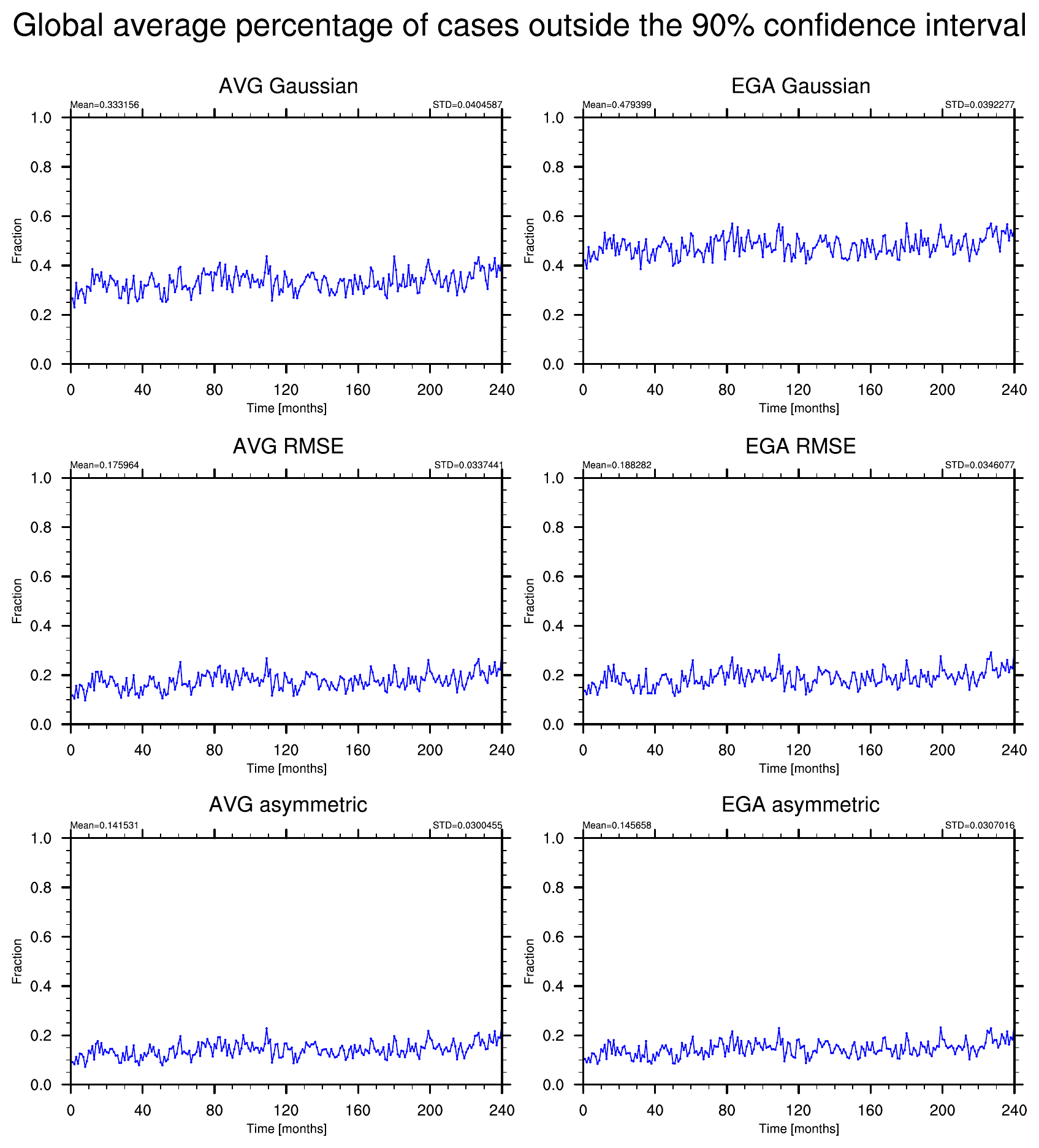} 
\caption{\label{fig:uas_ap=1_rel_year-crop} The fraction of the observations that were outside the predicted range of the $0.9$ confidence level for surface zonal wind. The three left panels show the fraction for the simple average, and the three right panels show it for the EGA forecaster. The two upper panels show $F_G$, the two center panels show $F_R$, and the two lower panels show $F_A$. The spatial average and the STD of the difference are denoted above each panel.}
\end{center}
\end{figure*}

Figures \ref{fig:tas_ap=1_rel_year-crop}-\ref{fig:uas_ap=1_rel_year-crop} show the fraction (in a spatial sense, i.e., for each time point, the fraction of grid cells) of the observations that were outside the predicted range of the $0.9$ confidence level. These figures show that the temporal variability of the reliability is small.

%\clearpage

%\clearpage

\subsection{ERA-interim reanalysis data}

The reliability of the uncalibrated and calibrated forecasts was also tested against the ERA-interim reanalysis data of two different resolutions. The reliability diagram based on the NCEP/NCAR reanalysis data was also compared with the reliability diagram obtained based on the ERA-interim reanalysis data (both the learning and the validation periods assumed the same reanalysis data) with the same spatial resolutions ($2.5^{\circ}X2.5^{\circ}$). We found only minor differences in the performances of the estimation methods when different reanalysis data were used. In addition, the difference between the reliability diagrams for the two ERA-interim resolutions ($0.75^{\circ}X0.75^{\circ}$ and $2.5^{\circ}X2.5^{\circ}$) was indiscernible.
The results show that the suggested method is not sensitive to the reanalysis data used. 

\begin{figure*}
\begin{center}
\includegraphics[width=24pc]{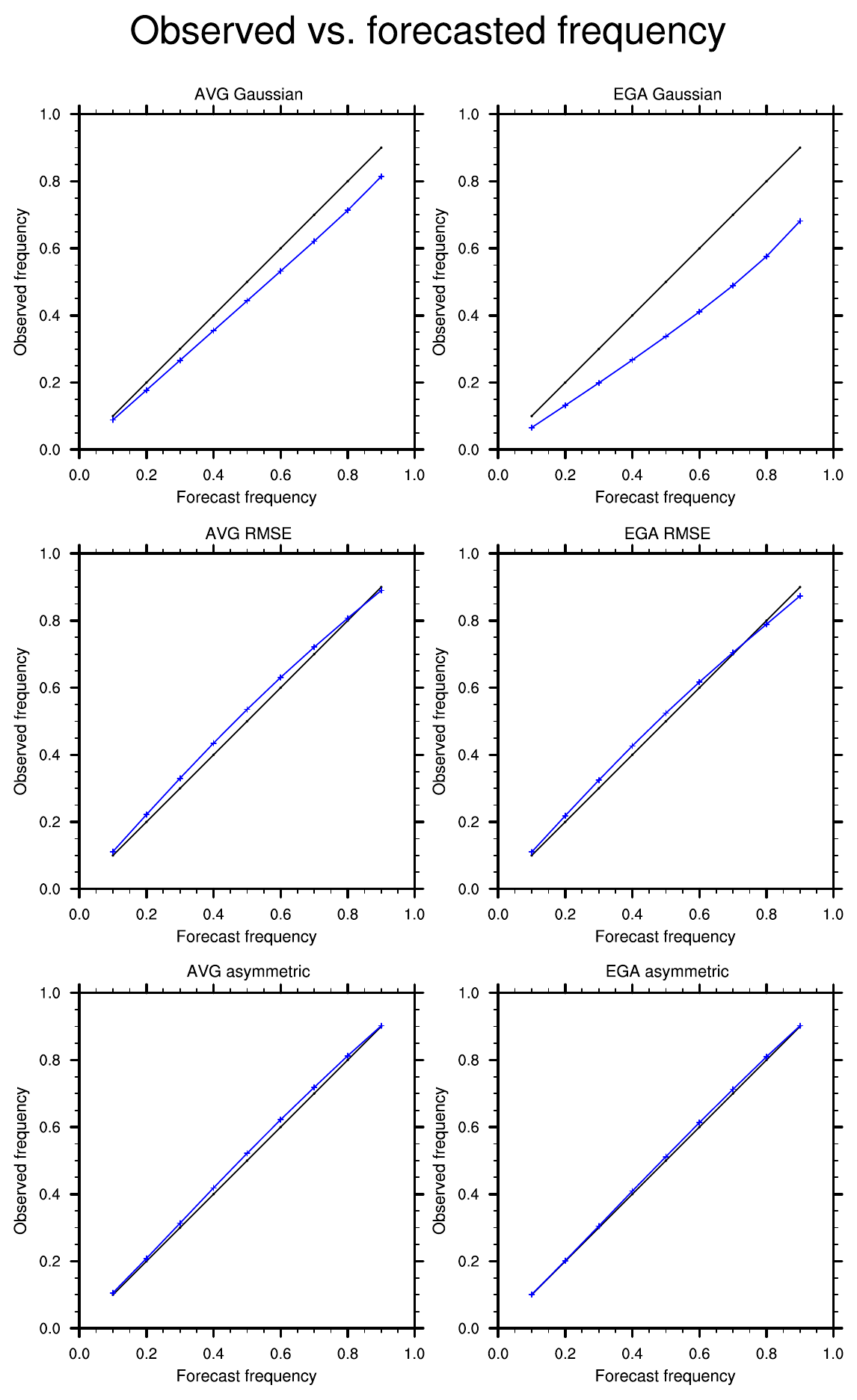} 
\caption{\label{fig:tas_ap=1_rel_diagram_ecmwf-crop} Reliability diagrams for the different estimation methods of the surface temperature uncertainty range. The observed frequency represents the spatial average over all the grid cells. The left panels show the reliability for the equally weighted ensemble, and the right panels show it for the EGA forecaster. The upper panels show the reliability of the Gaussian methods, the center panels show the reliability of the RMSE-corrected method, and the lower panels show the reliability of the asymmetric method. The source reanalysis data is the ERA-interim $2.5^{\circ}X2.5^{\circ}$}.
\end{center}
\end{figure*}

\begin{figure*}
\begin{center}
\includegraphics[width=24pc]{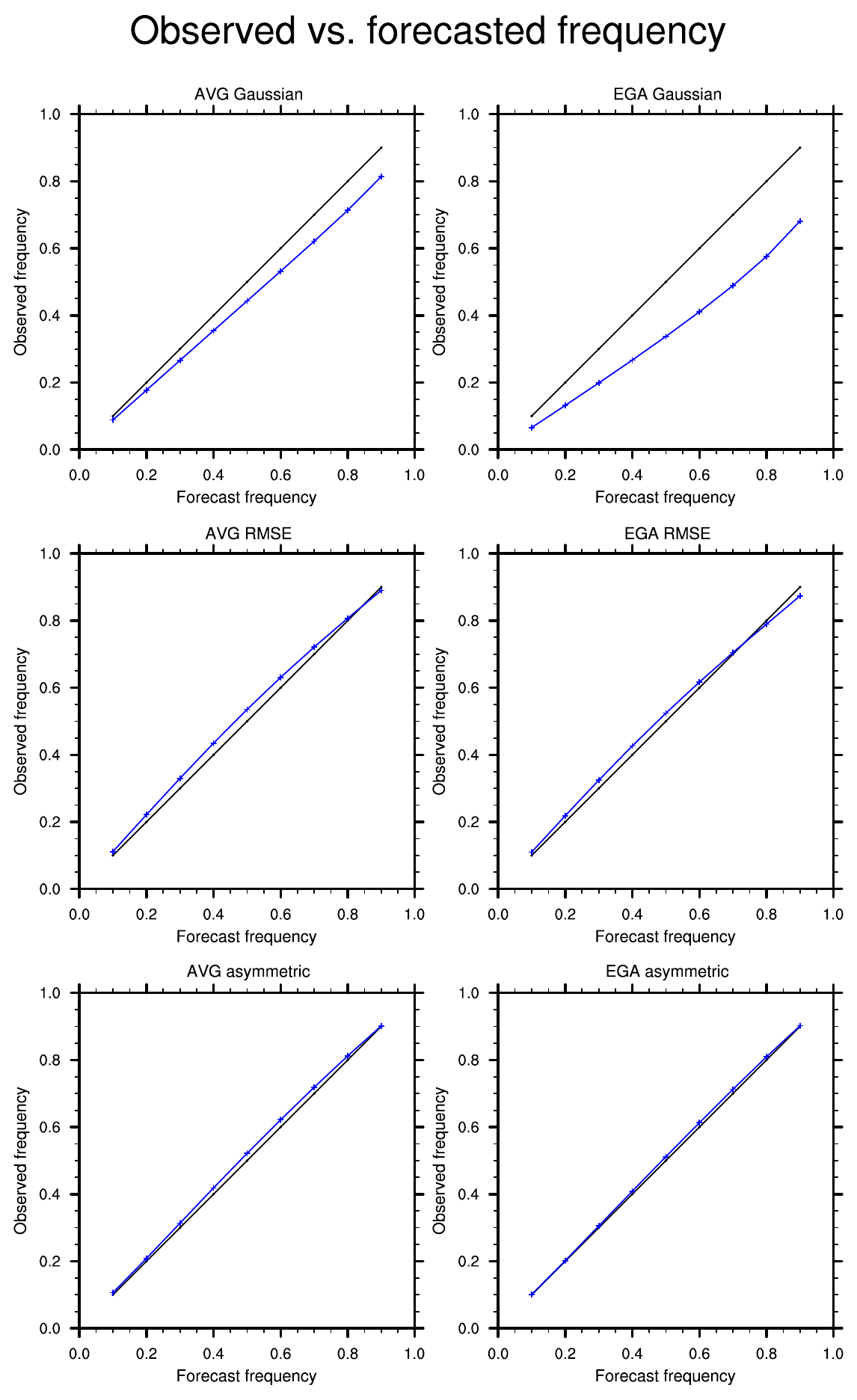} 
\caption{\label{fig:tas_ap=1_rel_diagram_ecmwf25-crop} Similar to Fig. \ref{fig:tas_ap=1_rel_diagram_ecmwf-crop} but for the ERA-interim $0.75^{\circ}X0.75^{\circ}$}.
\end{center}
\end{figure*}

%\clearpage

\subsection{Confidence interval}

\begin{figure*}
\begin{center}
\includegraphics[width=25pc]{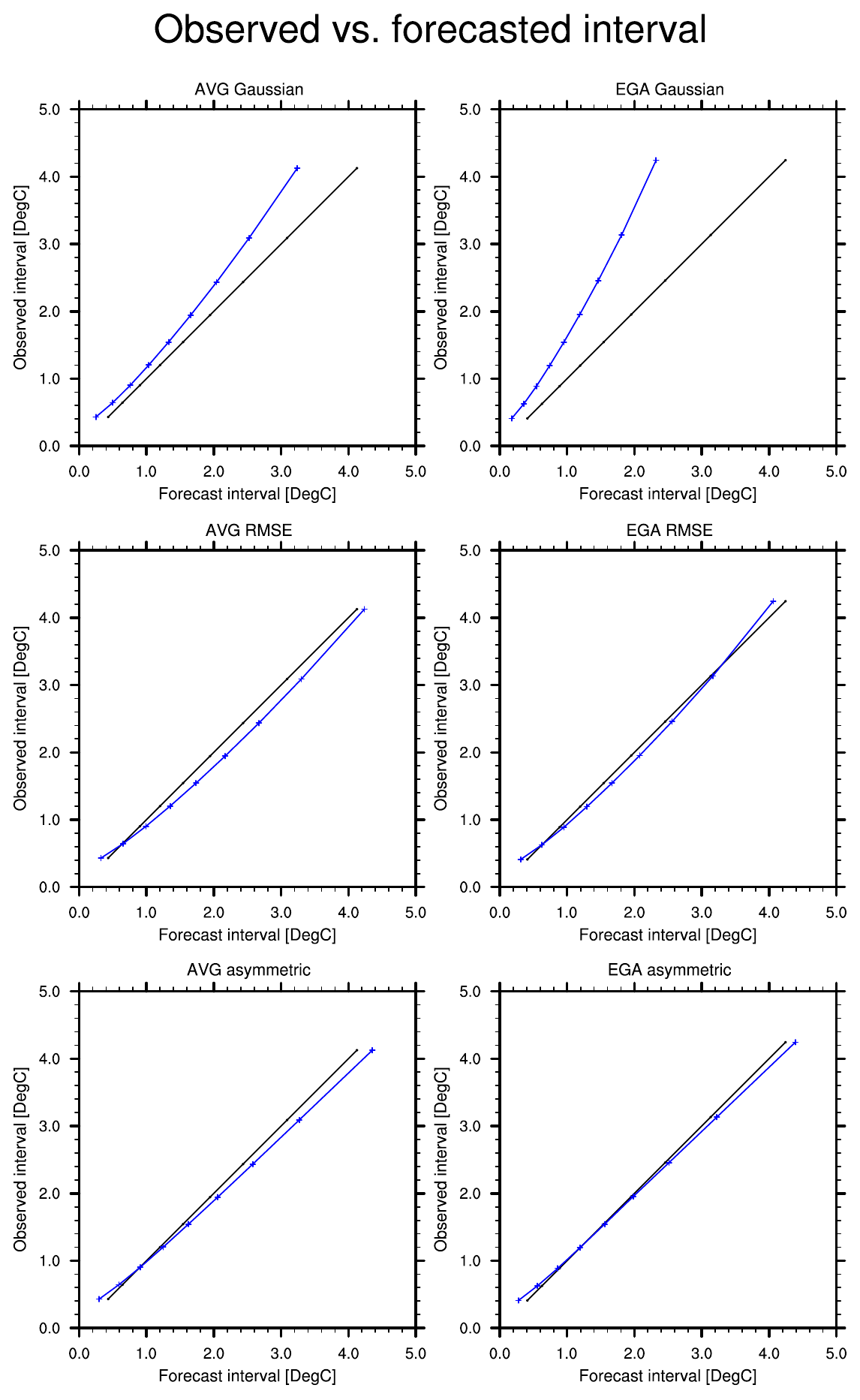} 
\caption{\label{fig:tas_ap=1_rel_interval-crop} The observed versus the predicted interval (in $\circ C$) for the surface temperature. Calculated from globally and temporally averaged intervals.}
\end{center}
\end{figure*}

\begin{figure*}
\begin{center}
\includegraphics[width=25pc]{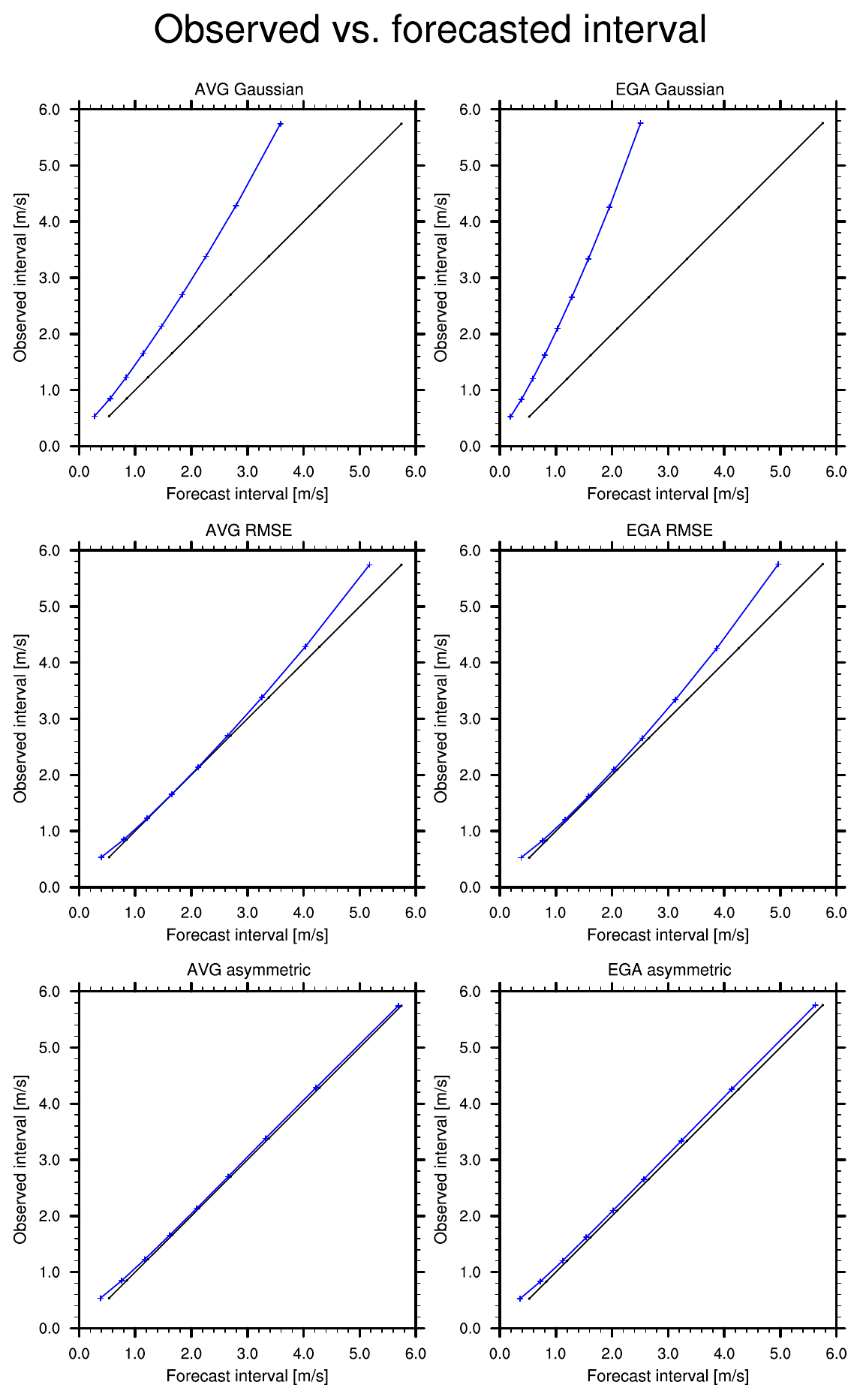} 
\caption{\label{fig:uas_ap=1_rel_interval-crop} The observed versus the predicted interval (in $m/s)$) for the surface zonal wind. Calculated from globally and temporally averaged intervals.}
\end{center}
\end{figure*}

Figures \ref{fig:tas_ap=1_rel_interval-crop}-\ref{fig:uas_ap=1_rel_interval-crop} show the predicted versus the observed confidence intervals.

%\clearpage

\end{document}